# The Duel: Strings versus Loops

*Physicists in search of the foundation of the world: how tiny objects can create matter, energy and even space and time – and possibly countless other universes*

By Rüdiger Vaas

---

An icy wind blows over the flat farmland close to the small German village of Golm, which was incorporated into the city of Potsdam in 2003. Outside Golm, in the middle of open fields, starts a new world. A spooky world it seems. For there fluctuate mysterious superposition of fabrics whose threads are supposed to be everything, and between which there is supposed to be nothing.

Strings vibrate in their highest tones, creating objects unknown to the surrounding people: photinos, gluinos, winos and zinos. But the familiar world is also a melody of strings. And it becomes stranger: allegedly, six or seven additional space dimensions exist curled up – in formal terms, compactified – like thin paper; higher-dimensional *branes* which flap like magic sheets in bizarrely formed Calabi-Yau spaces; moduli fields, mysterious anti-de Sitter spaces, a strange holographic principle which declares people to be holograms; unstable vacua, so that our universe could disappear with a whimper, and maybe even more universes than a person could possibly count – not only if he had to dedicate his entire lifetime, but also if he had the whole age of the universe at his disposal.

And all that, although these impressions would have surprised or disturbed a clueless hiker, belongs to the everyday business of a small but bustling group of researchers who met – steadfastly ignoring the rain soaked days outside – in the airily constructed building of the Max-Planck-Institute for Gravitational Physics (Albert-Einstein-Institute, AEI).

The atmosphere is cooperative but the occasion competitive. Although the conference theme sounds so conciliatory – "Strings Meet Loops" – it is debated keenly. At stake is literally everything: the universe and its foundation. Two well-developed theoretical approaches wrestle for a deeper basis of science. General relativity and quantum theory are the two undisputed pillars supporting the magnificent construction of contemporary physics, but on small space-time scales and at high energies they do not get along. For reconciliation, an even bolder theory is necessary. And on top of that this theory should – physicists being allowed to be immodest sometimes – unify the basic forces of nature and explain the riddles of the big bang and black holes.

String theory and quantum geometry (also known as loop quantum gravity) are the names of the two competitors. They follow the goal in different fashions and with different results. Hundreds of most gifted physicists worldwide have joined this project – and even invented new mathematical methods. For, in this unknown territory, there are no trails let alone a royal pathway. Thus the scientists have to fight their way through an exotic jungle – without any guarantee that they would not be hopelessly lost. Nobody can show concrete experimental proof yet. But many bizarre discoveries have already been made – and along the way quite a few precious fruits have been found. Nevertheless, string theorists and quantum geometers are not always friends.

"There is a certain reluctance to communicate in both camps, which we want to overcome. It would be better for everybody to think sometimes about problems from the others' perspective or to take their lessons to heart," says Hermann Nicolai, director of the AEI. To foster these communications and to inspect the successes and weaknesses was the main motivation for the meeting. Abhay Ashtekar, who organized the meeting with Nicolai, agrees and adds: "There have been many misconceptions about what the other side has assumed and achieved. So, we wanted the two communities to clarify the situation, set a language to communicate and, most importantly, learn from each other."

Ashtekar, professor of physics at the Pennsylvania State University and director of the Center for Gravitational Physics and Geometry located there, has developed quantum geometry together with Lee Smolin and Carlo Rovelli. According to this theory, space and time are not fundamental and independent of matter – they do not just represent a background metric as physicists say – but are themselves built from elementary objects called spin networks or sometimes loops. They form a fine mesh with "gaps" in between where literally nothing exists. Not even empty space, for space is only woven from the network and happens to look homogeneous from our blurred, distanced perspective – similar to pictures on these pages which on closer look do not appear smooth and connected but as patterns of individual pixels. Excitations of the spin networks – changeable states of lines and knots – build matter and energy in the universe. Also the forces of nature "live" on this weave. Time emerges, at least according to one idea, from those tiniest of the tiny changes in the spin network. Thus arises the picture of spin foams.

"A spin network is like a cross section through a spin foam, a kind of still picture," explains Robert Oeckl from the Universidad Nacional Autónoma de México. "Vice versa, we can understand a spin foam as the temporal evolution of a spin network. The smallest parts of a spin foam then correspond to space time atoms. Thus time would also be discrete. The 'beats' of time, certainly, are local – the universe does not evolve in simultaneous steps everywhere." String theory in comparison is much less radical. Oeckl: "In perturbative approaches to quantum gravity, and string theory is one example, the metric is decomposed into a background, which characterizes Euclidean space, and a second part which describes fluctuations of space. A big disadvantage is that there the causal structure is assumed to be static despite being changed dynamically by gravity."

This problem does not appear in quantum geometry – but at a price. "Imagine there is no space and time in the background; no canvas to paint the dynamics of the physical universe on. Imagine a play in which the stage joins the troupe of actors. Imagine a novel in which the book itself is a character," says Abhay Ashtekar, straining the imagination of his audience enormously. And he knows how to look at the bottom of things – i.e., how to glance so deeply into the structure of the world that in the glaring spotlight of the theory even space and time literally dissolve. "Yes, one can still do physics without sacrificing any mathematical precision."

Time and again Ashtekar returns to Albert Einstein whose legacy he continues, and whose theory of general relativity is to be extended. "In classical physics, Einstein taught us how to do this by weaving the gravitational field into the very fabric of space-time. In the resulting theory, general relativity, there is no background space-time, no inert arena, no spectators in the cosmic dance. Matter, through its gravity, tells space-time how to bend and curved space-time, in turn, tells matter how to move. However, classical physics is incomplete; it ignores the quantum world. Can we fuse the pristine, geometric world of Einstein's with quantum physics, without robbing it of its soul? Can we realize Einstein's vision at the quantum level?"

Still in the seventies the answer of every theorist would have been in the negative. But nowadays there are better reasons for finding a working theory of quantum gravity.

"Every great and deep difficulty bears in itself its own solution," Amitabha Sen quotes the physics Nobel laureate Niels Bohr. In 1981 Sen, as a student at the University of Chicago introduced a new mathematical description into physics, which allowed Ashtekar to express Einstein's general relativity in a new equivalent language – the basis for developing quantum geometry.

The central statement of this theory of quantum gravity according to Lee Smolin from the Perimeter Institute in Waterloo, Canada: "Space is not continuous but like atoms." This means that space has a discrete structure characterized by the Planck length $10^{-33}$ cm. "The smallest possible nonzero area is about a square Planck length, or $10^{-66}$ cm$^2$. The smallest nonzero volume is approximately a cubic Planck length, $10^{-99}$ cm$^3$. Thus, the theory predicts that there are about $10^{99}$ atoms of volume in every cubic centimeter of space. The quantum of volume is so tiny that there are more such quanta in a cubic centimeter than there are cubic centimeters in the visible universe ($10^{85}$)." Space atoms are made from the spin network. Smolin: "If we could draw a detailed picture of the quantum state of our universe – the geometry of its space, as curved and warped by the gravitation of galaxies and black holes and everything else – it would be a gargantuan spin network of unimaginable complexity, with approximately $10^{184}$ nodes."

Hermann Nicolai appears little impressed by quantum geometry. Beyond its highly developed mathematical formalism, he says, it is not yet completely clear if this theory can really transfer safely all the essential physical features of Einstein's theory, in particular its four dimensional covariance, to quantum theory. Furthermore, quantum geometry tries, first and foremost, to describe only gravity. The theory of matter is added by hand but not explained by deeper principles. "Here string theory is by far more ambitious because it not only wants to describe gravity correctly, but also explain the origin of matter," says Nicolai. Here the basic contradiction is that string theory says: Matter interactions cannot be ignored even in principle in order to make quantum gravity consistent. And the followers of quantum geometry: No, we simply try to transform Einstein's theory of relativity into a consistent quantum theory of gravity. Nicolai, who says of himself to have worked "on both sides of the fence," is not satisfied by this situation. "This contradiction has to be resolved at some stage." And he does not hold back his preference: "Personally, I tend toward the string side. There must be a reason for matter to exist in the world. If Einstein's theory would be consistent as quantum gravity one would have to ask: Why is there any matter at all?"

String theory claims to have the answer to this question. It understands elementary particles as excited states of tiny vibrating threads and can, in contrast to quantum geometry, describe all four forces of nature in a unified way. Its disadvantage is that it can be formulated only in nine or ten space dimensions. (Strictly speaking there are five different ten dimensional string theories and one eleven dimensional super gravity theory. All of them turned out to be closely related to one another and are nowadays seen only as boundary points of an encompassing, mostly unknown super theory christened M theory by the string theorist Edward Witten from the Institute of Advanced Study in Princeton.)

In view of its abstractness and exoticness the widespread popularity of string theory is astonishing. Last year the higher dimensional Calabi-Yau spaces even made it into the *New Yorker*, where Woody Allen humorously mocked them and other fuss of theoretical physics in a story of an office affair. Also, in the academic world the community of string theorists is at least ten times bigger than that of quantum geometry.

"String theory has more appeal within the physics community because it uses the standard language of background dependent quantum field theory. Since it is compatible with other areas of theoretical physics, many people were able to make a continuous transition from particle physics to string theory. The framework of loop quantum gravity, on the other hand, is new and different from anything else. Therefore, one needs to invest a lot of time to develop a new intuition," says Jerzy Lewandowski who gave a talk at the AEI about diverse "miracles" in the context of quantum geometry – theoretical breakthroughs and interrelations which enhanced his confidence in this approach. With Abhay Ashtekar, Thomas Thiemann and others, he contributed substantially to the proof of uniqueness and mathematical well-definedness of quantum geometry. In addition to that, Lewandowski emphasizes a different kind of unification in quantum geometry: "All forces 'live' in similar ways within the spin network and are treated alike in quantum geometry, although they are not unified in the same sense as in string theory."

Ashtekar suspects a further sociological reason for the numerical dominance of string theorists: There are simply many more particle physicists than relativists. "String theory naturally arose as a continuation of perturbative quantum field theory techniques which have been so successful in describing the other forces of nature. Quantum geometry arose from conceptual ideas that are at the heart of general relativity." And since, as one may speculate, many particle physicists were underoccupied following the fantastic confirmation of the standard model of matter and the subsequent wait for next generation accelerators to reach higher energies, they were forced to find themselves a new and spectacular area of research.

"Never in the history of physics has there been such an all-encompassing intellectual effort," says Nicolai about the string theorists. Compared to that the results are – except for much high calibre mathematics – certainly meagre. "The situation just shows that nature is still a bit more subtle than us."

Brian Greene who landed a best seller with his book *The Elegant Universe* (1999) spoke similarly in a recent interview with the *Scientific American*: "The universe in a sense guides us toward truths, because those truths are the things that govern what we see. String theory has been built up out of a lot of smaller ideas that a lot of people have contributed and been slowly stitching together into an ever more impressive theoretical edifice. But what idea sits at the top of the edifice, we still don't really know. When we do have that idea, I believe that it will be like a beacon shining down; it will illuminate the edifice, and it will also, I believe, give answers to critical questions that remain unresolved."

In the eyes of other scientists this is just wishful thinking, particularly since string theory with all its intellectual efforts has not really earned merits with physical applications and testable predictions – apparently it would not even be in contradiction with completely different data. Lee Smolin, for one, teases: "So far, no string theory background is known which is consistent with all features of the observed universe. They all have one or more of the following features, which each disagree with observation: no positive cosmological constant, unbroken supersymmetry, massless scalar fields."

"What's the good of a theory that can accommodate anything and the contrary of anything?" Carlo Rovelli from the University of Marseille has a sharp quantum geometry student saying to a string theory professor. "Our arguments have to be about the world we experience, not about a world made of paper." Exactly this sentence can be found already in the work *Dialog concerning the two chief world systems* (1632) of Galileo Galilei, which Carlo Rovelli took as the model for his also fictitious but biting article *A dialog on quantum gravity* (2003). There, the student and the professor continue the Galilean contest in a new guise. With much pleasure Rovelli dissects the weaknesses and unfulfilled promises of string theory. "The history of science is full of beautiful ideas that turned out to be wrong. In spite of the tremendous mental power of the people working in it, in spite of the string revolutions and the excitement and the hype, years go by and the theory isn't delivering physics. All the key problems remain wide open. The connection with reality becomes more and more remote."

That would not be Nicolai's judgment. "In fact I see some danger here. I do not believe, in particular, that string theory can survive twenty more years without a definitive and experimentally verifiable prediction." But this is a problem which quantum geometry also has to face. "The quantization of areas and volumes on the Planck scale will be even less verifiable than the supersymmetric particles in accelerator experiments, which are a consequence of string models."

Leonard Susskind brings this to the point: "Either all matter is strings, or string theory is wrong. This is one of the most exciting features of the theory." The professor of physics at Stanford University is one of the fathers of string theory, but certainly without being slavishly devoted to it. "String theory is either a theory of everything or a theory of nothing. The final evaluation of string theory will rest on its ability to explain the facts of nature, not on its own internal beauty and consistency. String theory is well into its fourth decade, but so far it has not produced a detailed model of elementary particles or a convincing explanation of any cosmological observation."

In 2003 Susskind tormented his colleagues with several unpleasant talks. For instance, many of the renowned cosmologists who met in Davis, California, to celebrate the new observations from the early universe, pulled a long face. Susskind shocked his audience with a number which gave pause to even those researchers who are familiar with large numbers from their everyday work. String theory would predict or need $10^{100}$ or even $10^{200}$ different vacuum states – at least. In other words: String theory has astronomically many physical solutions, and each of them could correspond to a universe with its own laws and constants of nature. If physicists are lucky, one among them is in accordance with our universe. But not even that is certain.

"The problem does not seem to be a lack of richness, but rather the opposite. String theory contains too many possibilities," summarized Susskind. "For most physicists, the ideal physical theory is one that is unique and perfect, in that it determines all that can be determined and that it could not logically be any other way. In other words, it is not only a theory of everything but it is the only theory of everything. To the orthodox string theorist, the goal is to discover the one true consistent version of the theory and then to demonstrate that the solution manifests the known laws of nature, such as the standard model of particle physics, with its empirical set of parameters."

Instead, string theorists stumbled upon a giant landscape they slyly call "stringscape." Yes, an enormously complex space of possibilities. "To mix metaphors: it is a stupendous haystack that contains googles of straws and only one needle. Worse still, the theory itself gives us no hint about how to pick among the possibilities."

At least, there is support for the stringscape from cosmologists who, in the context of the model of so-called eternal inflation, assume a multitude of different universes too. Susskind: "Thus it may be that the enormous number of possible vacuum solutions, which is the bane of particle physics, may be just what the doctor ordered for cosmology."

Michael Douglas from Rutgers University tries to make a virtue out of this necessity. At the "Strings Meet Loops" meeting he gave a talk via video-link, which was discussed with much controversy.

Douglas likewise takes at least $10^{200}$ different string vacuum states for granted. (Twenty years ago, even the number $10^{1500}$ was mentioned which, however, Douglas considers "out of the question." This prompted Hermann Nicolai to compare the situation with medieval theologists discussing the number of angels that can fit on the tip of a needle.) But he imagines certain selection principles would prevail allowing string theorists to pick the solutions interesting for our own universe, out of the many possible. Douglas even works on statistical methods to sift through the stringscape in search of prominent landmarks, and to scout out the structure of the stringscape in more detail.

To critics this effort appears futile as long as it is not even known whether any "realistic" universes are offered at all – or whether an infinite number of vacuum states indeed exist. Consequently, it would be impossible to verify the physics.

Susskind has invoked the so-called anthropic principle: We should not be surprised to live in our universe, for almost all others would not allow us to exist there – because there are no stars, for instance. Douglas is more cautious: "A valid physical theory must describe what we observe, but may or may not explain why we observe it. Anthropic arguments, while interesting, are not really central to fundamental physics, whose basic problem I would state as: Find the laws which describe all observations, and see if they can emerge from some complete and consistent framework."

A possible grasp into the hay stack to find the proverbial needle has been managed by a group of physicists around Shamit Kachru at Stanford. With some tricks they succeeded in finding a string vacuum which has at least a little resemblance to the vacuum state of our physical universe. Fernando Quevedo from Cambridge and his colleagues have in the meantime discovered more such "de Sitter spaces" in string theory which, so far, are not realistic but at least have a positive cosmological constant, a feature which appears necessary for the observed accelerated expansion of our universe. Certainly: If there are so many different vacuum states there is a danger of our own one existing only temporarily, being unstable in the long term, and decaying into a less energetic one. That would be the end of our universe. But Quevedo reassures: "All these transitions have a very large life-time – usually much bigger than the age of our universe."

"The awe for the math in string theory should not blind us. I think it is time to explore something else," is Rovelli's advertisement for quantum geometry. "The conventional mathematical formalism of quantum field theory relies very much on the existence of background space. There are therefore two possible strategies that we can adopt to construct a quantum theory of gravity. One is to undo Einstein's discovery and to reintroduce a fictitious background space. This is the strategy adopted by those who do not regard the general-relativistic revolution as fundamental, but as a sort of accident. And this is the strategy adopted in string theory. The second strategy is the one adopted by loop gravity: take general relativity seriously, directly face the problem that there is no background space in nature, and reconstruct quantum field theory from scratch in a form that does not require background space."

Abhay Ashtekar is more conciliatory: "Both communities agree that the final theory should be both background independent and should unify all interactions. The question is what one begins with, what one emphasizes to get the program going." Lee Smolin, who has also contributed to string theory, has a similar view: "Science works best when there is a variety of viewpoints investigated, and when there is room in the community for people who investigate a range of viable approaches to any unsolved problem."

Brian Greene likewise emphasizes common points: "My hope is that ultimately we're developing the same theory from different angles. It's far from impossible that we're going down our route to quantum gravity, they're going down their route to quantum gravity, and we're going to meet someplace. Because it turns out that many of their strengths are our weaknesses. Many of our strengths are their weaknesses."

"Before coming to a final judgment one has to see if and how all the pieces fit together in the end. For separate aspects one can always present some solutions," says Hermann Nicolai wrinkling his brow. Despite his preference for string theory he admits: "The aim is to find a theory which describes all of physics. Einstein already had the dream to condense everything into a single formula fitting on a sheet of paper. That is the dream of string theory, too – it just has not arrived there yet." Quantum geometry certainly does not provide that either. And it is also an open question whether one can succeed by joining forces. Nicolai: "I doubt that both approaches will finally mix harmoniously into a single theory since they start from diametrically opposite assumptions. It is not clear at all that in the end everybody will arrive at the same target. The purpose of our meeting was in fact to point out the differences and disputes."

It is certain that Einstein's dream has not come to its end yet. The great physicist called the two sides of his relativistic field equations – on one side geometry, on the other matter – "marble" and "wood." Following Einstein, Ashtekar and his colleagues also believe that in the end the world is built on the solid foundation of geometry rather than the brittle wood of matter. String theorists, on the other hand, can sound both tones with their vibrating strings. It remains to be seen who will toll the final sound – or if maybe even a third party is necessary.

---

**Acknowledgements:**
I am very grateful to Martin Bojowald, Michael Douglas, Jerzy Lewandowski, Hermann Nicolai, Robert Oeckl, Fernando Quevedo, Carlo Rovelli, Amitabha Sen, and Lee Smolin for their time, patience and great cooperation – and especially to Abhay Asthekar.

**Contact:**
Ruediger.Vaas@t-online.de

## Strings versus Loops: Comparing the Opponents

| properties and features | string theory | quantum geometry |
|---|---|---|
| fundamental objects | space, time, strings and branes | spin network or spin foam |
| number of spatial dimensions | 9 or 10 | 3 (more possible) |
| number of time dimensions | 1 | 1 |
| spacetime as a background metric | yes | no |
| modification of quantum theory and general relativity | yes | yes |
| conceptual unification of quantum theory and general relativity | no | yes |
| new physical principles necessary | yes | no |
| nature of matter | excitations of strings/branes | spin network states |
| explanation of the standard model of matter | in early research stage | not required, not possible |
| prediction of unknown elementary particles | yes | no |
| explanation of dark matter in the universe | possibly | no |
| explanation of dark energy in the universe | possibly | possibly |
| problems with infinities in the formalism | no (?) | no (?) |
| unification of quantization | not of space and time | yes |
| unification of interactions (forces of nature) | yes | no, but permitted |
| supersymmetry required and predicted | yes | no |
| uniqueness | no (many string vacua) | no (ambiguity of the Hamiltonian) |
| existence of many other universes | possible | unclear |
| explanation of black hole entropy | partly | yes |
| explanation of the big bang | possible (several) | possible |
| explanation of cosmic inflation | possible | possible |
| contact with low energy physics (everyday world) | in early research stage | in early research stage |
| description of scattering experiments | yes | not yet succeeded |
| testable predictions | in early research stage, partly falsified | in early research stage |

**Further reading:**

1. Brief introductions into quantum gravity:
http://www.damtp.cam.ac.uk/user/gr/public/qg_home.html
http://superstringtheory.com
http://www.qgravity.org

2. *Strings Meet Loops* conference website:
http://www.aei-potsdam.mpg.de/events/stringloop.html

3. *Welcome to Quantum Gravity*
Physics World (2003), vol 16, no. 11, pp. 21-22 and 27-47.
Special issue with articles by Fernando Quevedo, Leonard Susskind, Carlo Rovelli, and Giovanni Amelino-Camelia about string theory, quantum geometry and future experiments.

4. Woody Allen: *Strung Out*. The New Yorker (2003), July 28.
http://www.eps.org/aps/apsnews/1103/110319.html

5. Abhay Ashtekar's homepage:
http://cgpg.gravity.psu.edu/people/Ashtekar/

6. Abhay Ashtekar: *Addressing Challenges of Quantum Gravity Through Quantum Geometry: Black holes and Big-bang*. In: Daniel Iagolnitzer, Vincent Rivasseau, Jean Zinn-Justin (eds.): International Conference on Theoretical Physics. Ann. Henri Poincaré, vol. 4, suppl. 1, pp. 55-69. Birkhäuser, Basel 2003.
See also: http://arxiv.org/abs/math-ph/0202008

7. Brian Greene: *The Elegant Universe.* W. W. Norton, New York 1999.

8. Brian Greene, interview:
Scientific American (2003), vol. 289, no. 5, pp. 48-53.

9. Carlo Rovelli's homepage:
http://www.cpt.univ-mrs.fr/~rovelli/

10. Carlo Rovelli: *A dialog on quantum gravity*. International Journal of Modern Physics (2003), vol. D12, pp. 1509-1528.
http://arxiv.org/abs/hep-th/0310077

11. Lee Smolin: *Three Roads to Quantum Gravity*. Phoenix, London 2001.

12. Lee Smolin: *How far are we from the quantum theory of gravity?*
http://arxiv.org/abs/hep-th/0303185

13. Lee Smolin: *Atoms of Space and Time*. Scientific American (2004), vol. 290, no. 1, pp. 56-65.

14. Leonard Susskind: *A universe like no other*. New Scientist (2003), vol. 180, no. 2419, pp. 34-41.

15. Rüdiger Vaas: *Beyond Space and Time*.
http://arxiv.org/abs/physics/0401128
An introduction to quantum geometry. Translated from:
*Jenseits von Raum und Zeit.* bild der wissenschaft (2003), no. 12, pp. 50-56.

---